\documentstyle[12pt]{article}
\begin{document}
\begin{center}
{\bf B-L-violation in Softly Broken Supersymmetry 
           and Neutrinoless Double Beta Decay}
\bigskip 

{M. Hirsch, H.V. Klapdor-Kleingrothaus and S.G. Kovalenko$^*$ 
\bigskip

{\it Max-Planck-Institut f\"{u}r Kernphysik, P.O. 10 39 80, D-69029,
Heidelberg, Germany}
\bigskip

$^*$ {\it Joint Institute for Nuclear Research, Dubna, Russia}
}
\end{center}
\begin{abstract}
We prove a low-energy theorem valid for any model of weak scale softly
broken supersymmetry. It claims that the neutrino Majorana mass, 
the B-L violating mass of the sneutrino and the neutrinoless double beta 
decay amplitude are intimately related to each other such that if one of 
them is non-zero the other two are also non-zero and, vice versa, 
if one of them vanishes the other two vanish as well. The theorem is a 
consequence of the underlying supersymmetry and independent of the 
mechanisms of neutrinoless double beta decay and (s-)neutrino mass 
generation.
\end{abstract}
Neutrinos are special among the known fermions in the sense that - 
being electrically neutral - they could be either Dirac or Majorana 
particles. Experimentally at present only upper limits on neutrino 
masses have been firmly established, but there are also an 
accumulating number of hints for non-zero neutrino masses from, 
for example, the solar and atmospheric neutrinos (for a recent review 
see \cite{Smirnov}) as well as from recent LSND results \cite{LSND}. 
While the discovery of any non-zero neutrino mass would present a 
major breakthrough, unfortunately none of the above experiments 
could tell whether the neutrino is a Dirac or a Majorana particle. 

From a theoretical point of view Majorana neutrinos are clearly
preferred. In particular, Grand Unified Theories (GUTs) most naturally 
lead to Majorana neutrinos. A Majorana mass for the neutrinos could 
quite elegantly explain the observed smallness of neutrino masses via 
the see-saw mechanism \cite{see-saw}. Also various 1-loop contributions 
to the neutrino self-energy, allowed in extensions of the SM 
\cite{Lee84}-\cite{Zee}, induce a small Majorana mass for neutrinos. 
Nevertheless only experiments can finally settle the question about the 
nature of the neutrino.  
The importance of neutrinoless double beta ($0\nu\beta\beta$) decay 
derives from the fact that it is sensitive to the Majorana 
nature of neutrinos. 

That there is a generic relation between the amplitude of neutrinoless 
double beta ($0\nu\beta\beta$) decay and the (B-L)-violating 
Majorana mass 
of the neutrino has been recognized about 15 years ago \cite{SV}. 
A general theorem relating these two observables has been proven in 
\cite{SV}. It states that if any of the two quantities - the Majorana
neutrino mass or the neutrinoless double decay amplitude - vanishes 
the other one vanishes necessarily too and, vice versa, if one of them 
is non-zero the other one must also  
differ from zero. Recall, that  
$0\nu\beta\beta$-decay is strictly forbidden if the neutrino is a Dirac 
particle having only a (B-L)-conserving Dirac mass. 
This theorem is valid 
for any gauge model with spontaneously broken symmetry 
at the weak-scale, 
independent of the mechanism of $0\nu\beta\beta$-decay. The simple
neutrino exchange mechanism given in Fig.1(a) illustrates the 
above theorem explicitly. In this case the $0\nu\beta\beta$-decay 
amplitude is directly proportional to the small Majorana  mass 
$m^{\nu}_{M}$ of the neutrino.

Weak-scale softly broken supersymmetry implies new particles with 
masses of the order of $\sim M_W$  and new low-energy interactions. 
In view of this fact the $0\nu\beta\beta$-decay amplitude may 
non-trivially depend not only on the Majorana neutrino mass, as 
claimed by the above mentioned theorem \cite{SV}, but also on certain 
SUSY parameters. In SUSY models the neutrino ($\nu$) has a scalar 
superpartner the sneutrino ($\tilde\nu$.) Given that they are components 
of the same superfield there could be a certain interplay between the 
neutrino and sneutrino mass terms in a low-energy theory as a relic 
of the underlying supersymmetry. Such a relation 
indeed exists \cite{theorem1} and provides the basis for the present 
paper. 

In the present note we prove a low-energy theorem establishing an 
intimate relation between the neutrino Majorana mass, the 
(B-L)-violating sneutrino mass and the $0\nu\beta\beta$-decay amplitude. 
This theorem can be regarded as a SUSY generalization of the above 
cited theorem \cite{SV} proven for non-supersymmetric 
gauge theories. Our 
considerations use only the general structure of the low-energy 
effective Lagrangian assuming weak scale softly broken supersymmetry. 
The proof is based on symmetry arguments and thus generally valid, 
independent of 
mechanisms of neutrinoless double beta decay and (s-)neutrino mass 
generation. 

As shown in \cite{theorem1}, and as will also be demonstrated below, 
a self-consistent form of the neutrino and sneutrino mass terms is 
\begin{eqnarray} \label{complete}
{\cal L}^{\nu\tilde\nu}_{mass} = -\frac{1}{2} (m_M^{\nu}
\overline{\nu^c} \nu + h.c.)  - \frac{1}{2}(\tilde m_M^2
\tilde\nu_L \tilde\nu_L + h.c.) - 
\tilde m_D^2 \tilde\nu_L^* \tilde\nu_L.  
\end{eqnarray}
where $\nu = \nu^c$ is a Majorana field. The first two terms 
violate the global (B-L) symmetry while the last one respects
it. The first term is a Majorana mass term of the neutrino. 
We  call the second term a "Majorana"-like 
mass, while the third one is referred to as a "Dirac"-like  sneutrino 
mass term. This reflects an analogy with Majorana and Dirac mass terms
for neutrinos. The Dirac neutrino mass term 
$m_D^{\nu} (\bar\nu_L \nu_R + \bar\nu_R\nu_L)$ could also be included
in Eq. (\ref{complete}) but it is  not required by the self-consistency 
arguments. Note that $\tilde m_M^2$ is not a positively defined 
parameter. The further proof does not depend on the mechanism of mass 
generation in the low-energy theory. For the sake of simplicity and  
without any loss of generality we ignore possible neutrino mixing.  

The low-energy theorem we are going to prove consists of three 
statements. Two statements touch upon the set of three (B-L)-violating 
quantities: the neutrino Majorana  mass $m_M^{\nu}$, the "Majorana"-like 
sneutrino mass $\tilde m_M$ and the amplitude of 
$0\nu\beta\beta$-decay $R_{0\nu\beta\beta}$. The third statement 
relates this set of (B-L)-violating quantities to 
the (B-L)-conserving "Dirac"-like sneutrino mass $\tilde m_D$.

\underline{Statement 1}: If one of the three quantities 
$m_M^{\nu},\ \tilde m_M,\ R_{0\nu\beta\beta}$ vanishes, 
then the two others vanish, too. 
       
\underline{Statement 2} is an inverse to statement 1: 
If at least one of the three quantities $m_M^{\nu},\ \tilde m_M,$ or 
$R_{0\nu\beta\beta}$ is non-zero, then the two others are non-zero, too. 

\underline{Statement 3}: In the presence of $\tilde m_M^2\neq 0$
in Eq. (\ref{complete}) there must exist a "Dirac"-like (B-L)-conserving
sneutrino  mass term with $\tilde m_D^2\geq|\tilde m_M^2|$.

Let us turn to the proof of the first two statements. 
It is relatively easy to see that if at least one of the quantities 
is non-zero the two others are generated in higher orders of perturbation
theory as demonstrated in Fig.1, where only dominant diagrams are
shown. Internal lines in these diagrams are neutralinos $\chi_i$, 
gluinos $\tilde g$, charginos $\chi^{\pm}$, selectron $\tilde e$, 
u-squark $\tilde u$ and sneutrino $\tilde\nu$. 
The latter is to be identified with the B-L-violating 
``Majorana''  propagator proportional to $\tilde m_M^2$.
The sneutrino ``Majorana'' propagator was explicitly 
derived in ref. \cite{theorem1} and, for the sake of self-consistency 
of the current paper, is repeated below. 

The various diagrams lead to relations among the three (B-L)-violating 
observables, which we write down schematically
\begin{eqnarray} \label{1-loop}
%%%%%%%%%%%%%%%%%           
z_i &=& \sum_{i\neq j} a_{ij}\cdot z_j +  {\cal A}_i.
\end{eqnarray}
Here, $z_i$ can stand for $z_i = m_M^{\nu}$, $\tilde m_M^2$, 
$R_{0\nu\beta\beta}$. The coefficients $a_{ij}$ correspond to 
contributions of the diagrams in Fig.1(a)-(f) so that 
$i,j = a,b,c,d,e,f$.  
Terms ${\cal A}_i$ represent any other possible contributions. 
The explicit 
form of $a_{ij}$ and ${\cal A}_i$ is not essential in the following. 
Important is only the presence of a correlation between 
$m_M^{\nu}, \ \tilde m_M, \ R_{0\nu\beta\beta}$, expressed by 
eq. (\ref{1-loop}). 

Now we are going to prove that if $z_{i_1} = 0$, then  $z_{i_2} =
z_{i_3} = 0$ (the same will be true for any permutation). On the 
basis of Fig.(1) and Eqs. (\ref{1-loop}) one can expect such properties of 
the set of observables $z_i$.  Indeed $z_{i_1} = 0$ in the
left-hand side of Eq. (\ref{1-loop}) strongly disfavors $z_{j_2}\neq 0$
and $z_{j_3}\neq 0$, because it requires either all the three terms 
in the right-hand sides to vanish or their net cancelation. 
The latter is "unnatural". Even if such a cancelation would be done by 
hand, using (unnatural) fine-tuning 
of certain parameters, in some specific order of perturbation theory, 
it would be spoiled again in higher orders of perturbation theory. 
The cancelation of all terms in the right-hand 
side of Eqs. (\ref{1-loop}) 
in all orders of perturbation theory  could only be 
guaranteed by a special unbroken symmetry. 
Let us envisage this possibility in details.

The effective Lagrangian of a generic model of weak scale softly broken 
supersymmetry contains after electro-weak 
symmetry breaking the following 
terms \cite{Haber}
\begin{eqnarray} \label{L1}
{\cal L} &=& - \sqrt{2} g \epsilon_i \cdot 
\overline{\nu}_L\chi_i\tilde\nu_L   
- g \epsilon_i^- \cdot \overline{e}_L\chi_i^-\tilde\nu_L   
- g \epsilon_i^+ \cdot \overline{\nu}_L\chi_i^+\tilde e_L + \\ \nonumber
&+&  \frac{g}{\sqrt{2}} (\overline{\nu}_L \gamma^{\mu} e_L + 
\overline{u}_L \gamma^{\mu} d_L)W^+_{\mu}
+ g \cdot \bar\chi_i\gamma^{\mu} (O^L_{ij}P_L +
                                  O^R_{ij}P_R)\chi^+_j W^-_{\mu} +  
                                                             ... + h.c. 
\end{eqnarray}
Dots denote other terms which are not essential 
for further consideration.  
Here, $\tilde\nu_L$ and $\tilde e_L$ represent scalar superpartners of 
the left-handed neutrino $\nu_L$ and electron $e_L$ fields. The 
chargino $\chi^{\pm}_i$ and neutralino $\chi_i$ are superpositions of 
the gaugino and the higgsino fields. 
The contents of these superpositions 
depends on the model. Note that the neutralino is a Majorana field 
$\chi_i^c = \chi_i$. The explicit form of the coefficients 
$\epsilon_i, \ \epsilon^{\pm}_i $ and $O^{L,R}_{ij} $ 
is also unessential. 
For the case of the MSSM one can  find them, 
for instance in \cite{Haber}. 
Eq. (\ref{L1}) is a general consequence of the underlying weak scale 
softly broken supersymmetry  and the spontaneously broken electro-weak
gauge symmetry. 

The Lagrangian (\ref{L1}) does not posses any continuous symmetry having 
non-trivial B-L transformation properties. Recall, that $U(1)_{B-L}$
is assumed to be broken since we admit B-L-violating mass terms in
Eq. (\ref{complete}). However, there might be an appropriate unbroken 
discrete symmetry. Let us specify this discrete symmetry 
group by the following field transformations 
\begin{eqnarray} \label{discrete}
\nu &\rightarrow & \eta_{\nu} \nu, \ \ \
\tilde\nu\rightarrow\eta_{\tilde\nu} \tilde\nu, \ \ \
e_L \rightarrow  \eta_{e} e_L, \ \ \ 
\tilde e_L  \rightarrow \eta_{\tilde e} \tilde e_L, \\  \nonumber
W^+ & \rightarrow & \eta_{_W} W^+,  \ \ \    
\chi_{i}\rightarrow\eta_{\chi_{i}} \chi_{i},\ \ \ 
\chi^+  \rightarrow  \eta_{\chi^+} \chi^+.
\end{eqnarray}
Here $\eta_i$ are phase factors. 
Since the Lagrangian (\ref{L1}) is assumed to be invariant under these
transformations one obtains the following relations
\begin{eqnarray} \label{symm_rel}
\eta_{\nu}^* \eta_{\tilde\nu} \eta_{\chi_i} &=& 1, \ \ \  
\eta_{e} \eta_{\chi^+}\eta_{\tilde\nu}^* = 1, \ \ \ ...\\ \nonumber
\eta_{e} \eta_{_W} \eta_{\nu}^* &=& 1, \ \ \ 
\eta_W^* \eta_{\chi^+} \eta_{\chi_i}^* = 1, \ \ \ ....  
\end{eqnarray}
Dots denote other relations which are not essential here.
The complete set of these equations defines the admissible 
discrete symmetry group of the Lagrangian in Eq. (\ref{L1}).

Let us find the transformation property of the operator structure 
responsible for $0\nu\beta\beta$-decay under this group. 
At the quark level $0\nu\beta\beta$-decay implies the transition 
$dd\rightarrow uuee$,  described by the effective operator
\begin{eqnarray} \label{operator}
{\cal O}_{0\nu\beta\beta} = \alpha_i\cdot \bar u 
\Gamma_i^{(1)} d\cdot \bar u \Gamma_i^{(1)} d\cdot 
\bar e \Gamma_i^{(2)} e^c,
\end{eqnarray} 
where $\alpha_i$ are numerical constants,  $\Gamma_i^{(k)}$ are 
certain combinations of Dirac gamma matrices. The
$0\nu\beta\beta$-decay amplitude $R_{0\nu\beta\beta}$ is related to
the matrix element of this operator 
\begin{eqnarray} \label{ampl}
R_{0\nu\beta\beta} \sim 
<2 e^- (A, Z+2)| {\cal O}_{0\nu\beta\beta} |(A,Z)>
\end{eqnarray}
where $(A,Z)$ is a nucleus with the atomic weight $A$ and the total 
charge $Z$. The operator in Eq. (\ref{operator}) transforms under 
the group (\ref{discrete}) as follows
\begin{eqnarray} \label{operator_trans}
{\cal O}_{0\nu\beta\beta}\rightarrow
\eta_{0\nu\beta\beta}{\cal O}_{0\nu\beta\beta} 
\end{eqnarray} 
 with 
\begin{eqnarray} \label{eta_dbd}
\eta_{0\nu\beta\beta} =  \eta_d^*\eta_u\eta_e
\end{eqnarray}

 Solving Eqs. (\ref{symm_rel}), (\ref{eta_dbd}), one finds
\begin{eqnarray} \label{sol}
\eta_{\nu}^2 = \eta_{\tilde\nu}^2 = \eta_{0\nu\beta\beta}^2.
\end{eqnarray}
This relation proves the statements 1,2.  To see this we note 
that the observable quantity 
$z_i = (m_M^{\nu}, \ \tilde m_M, \ R_{0\nu\beta\beta})$
is forbidden by this symmetry if the corresponding discrete group
factor is non-trivial, {\it i.e.}  $\eta_i^2\neq 1$. Contrary, if  
$\eta_i^2= 1$, this quantity is not protected by the symmetry 
and appears in higher orders of perturbation theory, even if 
it is not included at the tree-level.  Relation (\ref{sol})
claims that if one of the $z_i$ is forbidden then the two others 
are  also forbidden and, vice versa, if one of them is not forbidden 
they are all not forbidden. Thus, statements 1,2 are proven.

One can derive the following interesting corollary from statements 1,2.
                            
{\it Corollary 1}:             
 If the (B-L)-violating sneutrino "Majorana" mass term is for certain 
reason absent in the low-energy theory, {\it i.e.} $\tilde m_M^2=0$, 
then the neutrino has no Majorana mass and 
neutrinoless double beta decay is forbidden. 

Now let us turn to the statement 3. 
Consider the last two terms of Eq. (\ref{complete}) which we denote as 
${\cal L}^{\tilde\nu}_{mass}$ and use the real field representation for 
the  complex scalar sneutrino field
$\tilde\nu = (\tilde\nu_1 + i \tilde\nu_2)/\sqrt{2}$, where 
$\tilde\nu_{1,2}$ are real fields. Then
\begin{eqnarray} \label{snu_DM}
{\cal L}^{\tilde\nu}_{mass} =  - \frac{1}{2}(\tilde m_M^2
\tilde\nu_L \tilde\nu_L + h.c.) - \tilde m_D^2 \tilde\nu_L^*
\tilde\nu_L   = - \frac{1}{2}  \tilde m_1^2
\tilde\nu_1^2 -  \frac{1}{2}  \tilde m_2^2 \tilde\nu_2^2 
\end{eqnarray}
where $\tilde m_{1,2}^2 = \tilde m_D^2 \pm |\tilde m_M^2|$. Assume the
vacuum state is stable. Then $\tilde m_{1,2}^2\geq 0$, i.e. 
$\tilde m_D^2 \geq |\tilde m_M^2|$, otherwise the vacuum is unstable and
subsequent spontaneous symmetry breaking occurs via non-zero 
vacuum expectation values of the sneutrino fields $<\tilde\nu_i>\neq 0$. 
The broken symmetry in this case is the R-parity. It is a discrete
symmetry defined as $R_p = (-1)^{3B+L+2S}$, where $S,\ B$ and $L$
are the spin, the baryon and the lepton quantum number.

This completes the proof of the theorem consisting of the above-given 
three statements, and moreover shows that a self-consistent
structure of mass terms of the neutrino-sneutrino sector is given by
Eq. (\ref{complete}).

Let us finally present the explicit form of the above mentioned 
(B-L)-violating ``Majorana'' propagator $\Delta_{\tilde\nu}^M$ for 
the sneutrino  \cite{theorem1}. It can be 
derived by the use of the real field representation 
as in eq. (\ref{snu_DM}). 
For comparison we also give the (B-L)-conserving ``Dirac''
$\Delta_{\tilde\nu}^D $  sneutrino propagator,
\begin{eqnarray} \label{propagators_1}
\Delta_{\tilde\nu}^D(x-y) &=&
-\frac{i}{2}(\Delta_{\tilde m_1}(x-y) + \Delta_{\tilde m_2 }(x-y)),  \\
\Delta_{\tilde\nu}^M(x-y)= 
&=& -\frac{i}{2}(\Delta_{\tilde m_1}(x-y) - \Delta_{\tilde m_2 }(x-y)),  
\end{eqnarray}
where
\begin{eqnarray} \label{def_1}
\Delta_{\tilde m_i}(x) = \int\frac{d^4 k}{(2\pi)^4}
\frac{e^{-ikx}}{\tilde m_i^2 - k^2 - i\epsilon}
\end{eqnarray}
is the ordinary propagator for a scalar particle with mass $\tilde m_i$. 
Using the definition of $\tilde m_{1,2}$ as in Eq. (\ref{snu_DM}) 
one finds    
\begin{eqnarray} \label{propagators_2} 
\Delta_{\tilde\nu}^D(x) &=& \int\frac{d^4 k}{(2\pi)^4}
\frac{\tilde m_D^2 - k^2}{(\tilde m_1^2 - k^2 - i\epsilon)
(\tilde m_2^2 - k^2 - i\epsilon)} e^{-ikx}, \\
\Delta_{\tilde\nu}^M(x) &=& -  \tilde m_M^2 \int\frac{d^4 k}{(2\pi)^4}
\frac{e^{-ikx}}{(\tilde m_1^2 - k^2 - i\epsilon)
(\tilde m_2^2 - k^2 - i\epsilon)}. 
\end{eqnarray}
It is seen that in absence of 
the (B-L)-violating sneutrino ``Majorana''-like 
mass term $\tilde m_M^2 =0$ the (B-L)-violating propagator
$\Delta_{\tilde\nu}^M$  vanishes while 
the (B-L)-conserving one $\Delta_{\tilde\nu}^D$  becomes the ordinary 
propagator of a scalar particle
with mass $\tilde m_{1} = \tilde m_{2} = \tilde m_D$. 
According to Eq. (\ref{snu_DM}) the parameter $\tilde m_M^2$ describes a
splitting in the sneutrino mass spectrum. This mass splitting
parameter can be probed by searching for (B-L)-violating exotic
processes such as neutrinoless double beta decay as discussed in 
the present note. It is obvious from Eq. (\ref{1-loop}) corresponding 
to the diagram in Fig.1(f) that certain constraints on $\tilde m_M^2 $
can also be obtained from the experimental upper bound on the neutrino
mass. Probably, $\tilde m_M^2$ can also be constrained by  accelerator 
searches for supersymmetry.  However this possibility might require 
unrealistic energy resolution for detectors if the above mentioned
$0\nu\beta\beta$-decay and/or $m_{\nu}$ constraints on $\tilde m_M^2$
turned out to be too stringent. 
We are going to analyze these questions in a separate paper.

In summary, we have proven a low-energy theorem for weak scale softly
broken supersymmetry relating the (B-L)-violating mass terms of the
neutrino and the sneutrino as well as the amplitude of neutrinoless
double beta decay. This theorem can be considered as a supersymmetric 
generalization of the well know theorem \cite{SV} relating only 
neutrino Majorana mass and the neutrinoless double beta decay amplitude.

\bigskip
\centerline{\bf ACKNOWLEDGMENTS}

We thank V.A.~Bednyakov,  for helpful discussions.
The research described in this publication was made possible in part 
(S.G.K.) by Grant GNTP 315NUCLON from the Russian ministry of science. 
M.H. would like to thank the Deutsche Forschungsgemeinschaft
for financial support by grants kl 253/8-2 and 446 JAP-113/101/0.

{\large\bf Figure Captions}\\

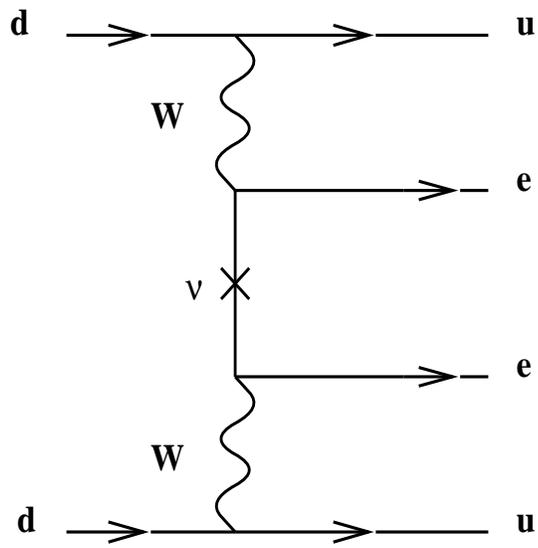
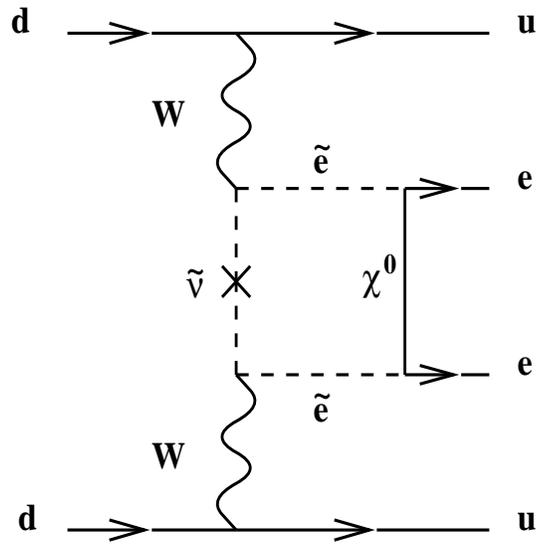
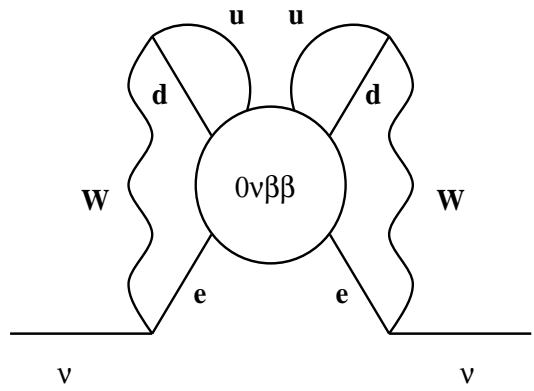
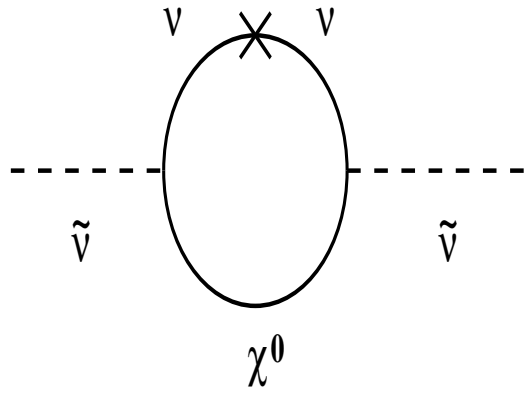
Fig.1.: Lowest order perturbation theory diagrams representing the 
relation between the neutrino Majorana mass $m_M^{\nu}$, the 
"Majorana"-like (B-L)-violating sneutrino mass $\tilde m_M$ and the 
amplitude of neutrinoless double beta decay $R_{0\nu\beta\beta}$. 
(a) the neutrino contribution and (b) an example of 
sneutrino contribution to the $0\nu\beta\beta$-decay amplitude 
$R_{0\nu\beta\beta}$. 
$0\nu\beta\beta$-vertex contribution to (c) the neutrino Majorana mass 
and (d) to the "Majorana"-like sneutrino mass; 
(e) neutrino contribution to the sneutrino "Majorana"-like mass 
and (f) sneutrino contribution to the neutrino Majorana mass. 
Crossed (s)neutrino lines correspond to the B-L-violating propagators.
        

\begin{thebibliography}{99}
%1
%%%%%%%%%%%%%%%%%%%%%%%%%%%%%%%%%%%%%%%%%%%%%%%%%%%%%%%%%%%%%%%%%%%%%
\bibitem{Smirnov} A.Yu. Smirnov, Plenary talk given at 
{\it 28th International Conference on High energy physics,} 25 - 31
   July 1996, Warsaw, Poland;  hep-ph/9611465
%
Y. Suzuki, Plenary talk at the same {\it Conference}.
% 
\bibitem{LSND} The LSND Collaboration, C. Athanassopoulos 
{\it et al.}, Phys.Rev.Lett. {\bf 77},  3082  (1996); {\it ibid.}   
{\bf 75}, 2650 (1995).  
%
%%%%%%%%%%%%%%%%%%%% theorem %%%%%%%%%%%%%%%%%%%%%%%%%%%%%%%%%%%%%%%%
%
\bibitem{SV} J. Schechter and J.W.F. Valle, 
             Phys.Rev. D {\bf 25}, 2951 (1982);
%
            J.F. Nieves, Phys.Lett. B {\bf 147}, 375 (1984);
            E. Takasugi, Phys.Lett. B {\bf 149}, 372 (1984);
            B. Kayser, in Proc. of {\it the XXIII Int. Conf on High
            Energy Physics,} ed. S. Loken (World Scientific
            Singapore, 1987), p. 945;
            S. Petcov, in Proc. of {\it 86' Massive Neutrinos in 
            Astrophysics and in Particle Physics,} ed. O. Fackler and
            J. Tr\^an Than V\^an (Editions Frontieres,
            Gif-sur-Yvette, France, 1986), p. 187;
            S.P. Rosen, UTAPHY-HEP-4 and hep-ph/9210202.
%
%%%%%%%%%%%%%%%%%%%%%%%%%%%%%%%%%%%%%%%%%%%%%%%%%%%%%%%%%%%%%%%%%%%%%
\bibitem{see-saw} M. Gell-Mann, P. Ramond, and R. Slansky, 
in  {\it Supergravity}, ed. F. van  Nieuwenhuizen and  D.Freedman,
(North Holland, Amsterdam, 1979), p.315; T. Yanagida, {\it Proc. of
the Workshop on Unified Theory and Baryon Number of the Universe},
KEK, Japan, 1979; S. Weinberg, Phys.Rev.Lett. {\bf 43}, 1566 (1979). 
%
%%%%%%%%%%%%%%%%%%%%%%%%%%%%%%%%%%%%%%%%%%%%%%%%%%%%%%%%%%%%%%%%
%%%%     1-loop contrib. to neutrino mass               %%%%%%%%
%%%%%%%%%%%%%%%%%%%%%%%%%%%%%%%%%%%%%%%%%%%%%%%%%%%%%%%%%%%%%%%% 
%
\bibitem{Lee84} I. H. Lee,  Nucl. Phys. B {\bf 246},  120 (1984);
                                   Phys.Lett. B {\bf 138},  121 (1984).
%
\bibitem{mnv85} R. Mohapatra, S. Nussinov and J.W.F. Valle,
                        Phys.Lett. B {\bf 165},  417 (1985).
%
\bibitem{barb90}   R. Barbieri et al. Phys. Lett. B {\bf 238},  
                   86  (1990).
%9
\bibitem{hs84} L. Hall and M. Suzuki, 
                 Nucl.Phys. B {\bf 231},  419 (1984).
%10
\bibitem{mn90} E. Ma and D. Ng, Phys. Rev. D {\bf 41}, 1005  (1990),
% 
\bibitem{Zee} A. Zee, Phys.Lett. B {\bf 93},  389 (1980).
%
%%%%%%%%%%%%%%%%%%%%%%%%%%%%%%%%%%%%%%%%%%%%%%%%%%%%%%%%%%%%%%
%
\bibitem{theorem1}  M. Hirsch, H.V. Klapdor-Kleingrothaus and
                    S.G. Kovalenko, hep-ph/9701253. 
%
%%%%%%%%%%%%%%%%%%%%%%%%%%%%%%%%%%%%%%%%%%%%%%%%%%%%%%%%%%%%%%%%%
%
\bibitem{Haber} H.E. Haber and G.L.Kane, 
                Phys.Rep. {\bf 117},  75  (1985);
                H.P. Nilles, Phys.Report. {\bf 110},  1 (1984).
% 
\end{thebibliography}
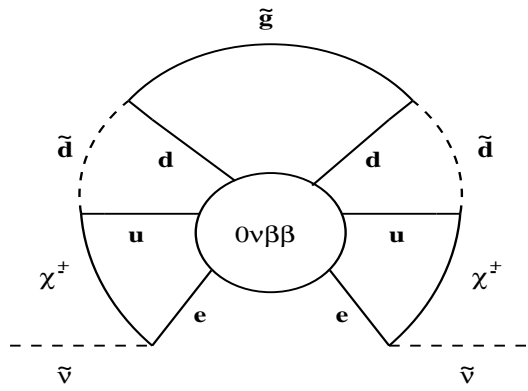
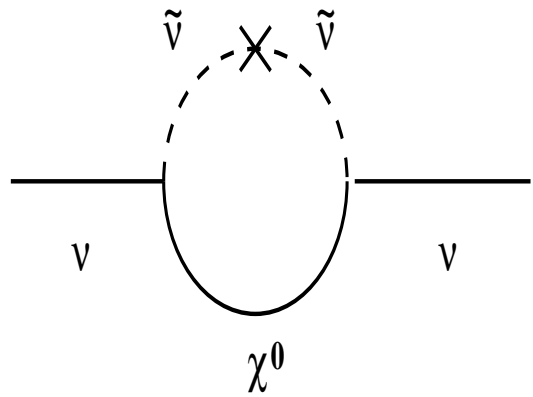
\end{document}